\documentclass[twocolumn,fleqn]{svjour2}    
\smartqed  
\usepackage{graphicx}
\begin{document}

\title{ Path-integral Monte Carlo study of phonons in the bcc phase of $^3$He
}


\author{ V.Sorkin  \and
        E. Polturak \and
        Joan Adler
}


\institute{  V.Sorkin \at
              Physics Department, Technion - Israel Institute of Technology, Haifa,
              Israel, 32000 \\
              Tel.: +972-4-8292043\\
              Fax:  +972-4-8295755\\
              \email{phsorkin@technunix.technion.ac.il}           
           \and
              E. Polturak \at
              Physics Department, Technion - Israel Institute of Technology, Haifa,
              Israel, 32000 \\
              Tel.: +972-4-829-2027\\
              Fax:  +972-4-8295755\\
              \email{emilp@physics.technion.ac.il} 
             \and
              Joan Adler \at
              Physics Department, Technion - Israel Institute of Technology, Haifa,
              Israel, 32000 \\
              Tel.:+972-4 829-3937\\
              Fax: +972-4-8295755\\
              \email{phr76ja@tx.technion.ac.il} 
}

\date{Received: 1.03.2006 / Accepted: date}

\maketitle

\begin{abstract}

                Using  Path Integral Monte Carlo and the Maximum Entropy method, we calculate
                the dynamic structure factor of solid $^3$He in the bcc phase at a finite temperature
                 of T = 1.6 K and a molar volume of 21.5 cm$^3$. 
		 From the single phonon dynamic structure
                 factor, we obtain both the longitudinal and transverse
                 phonon branches  along the main crystalline directions, [001], [011] and
                 [111]. Our results are compared with other theoretical
                 predictions and available experimental data.



\end{abstract}

\section{Introduction}
\label{intro}
The two isotopes of solid helium, $^3$He and $^4$He, are the most prominent examples of
quantum solids~\cite{Glyde,Dobbs}.
These isotopes display highly anharmonic dynamics:
their atoms are strongly correlated,
loosely bound and make large excursions to the nearest neighbor sites.
The effects are especially significant in the bcc phase of solid helium.

The dynamics of solid $^4$He, in particular the phonon spectrum,
has been studied over years in inelastic neutron scattering experiments~\cite{Osgood,Osgood1,Osgood2,Emil}.
In contrast to $^4$He, there are no experimental measurements of phonon spectra of $^3$He
(in the bcc phase) due to the large neutron  absorption cross-section
of this isotope.~\cite{Glyde,Senesi} Recently, however,
the use of inelastic X-ray scattering~\cite{Seyfert,Burkel} offers the possibility of
experimentally determining the phonon branches of $^3$He.
In addition, new neutron scattering experiments are being planned,~\cite{Schottl,Schanen}
which will be able to acquire data for a sufficiently long time
 to observe inelastic scattering. The  present study
has been motivated by these developments.

Strong anharmonic effects and atomic correlations
make  the theoretical calculation of phonon spectra of $^3$He
very difficult.  Several theoretical methods have been developed to treat this
problem: Glyde and Khanna devised a self-consistent phonon t-matrix formalism~\cite{Glyde_ph},
Horner applied a many-body perturbation technique~\cite{Horner,Horner_ph}, and
Koehler and Werthamer employed a variational approach.
These theoretical calculations are based on a variational perturbative theory
and implemented at zero-temperature.\cite{Glyde}
The predictions of these models are in quantitative disagreement with each other,
due to differences in the effective potentials used to evaluate phonon interactions
and details of the numerical methods employed.~\cite{Koehler}

As a complementary approach to these methods
we decided to study the excitations in bcc solid helium
$^3$He, by performing Quantum Monte Carlo  simulations at a
finite temperature.  We use Path Integral Monte Carlo
(PIMC)\cite{Ceperley_RMP}, which is a non-perturbative
numerical method, that allows, in principle, simulations of
quantum systems without any assumptions beyond the Schr$ \rm \ddot
o$dinger equation. The two body interatomic He-He potential
\cite{Aziz} is the only input for the PIMC simulations. In our
study the Universal Path Integral code of Ceperley\cite{Ceperley_RMP}
 was adapted to calculate the phonon
branches at finite temperature.
The PIMC method, in conjunction with Maximum Entropy (MaxEnt) techniques,~\cite{MaxEnt} was
employed to calculate the phonon spectra of liquid~\cite{spect}
and solid~\cite{Sorkin} helium ($^4$He) at finite temperatures.
The calculated phonon spectra are in good agreement with
the spectra measured by inelastic neutron scattering experiments.

The results of our study include the numerical calculation of all
the phonon branches of bcc $^3$He at a molar volume 21.5 cm$^3$.
Details of our simulations are
described in Sec. 2, the results of the calculations are presented in Sec. 3.

\section{ Method}

The PIMC method used in our simulations is  based on the formulation
of quantum mechanics in terms of path integrals. It has been described in
detail by Ceperley ~\cite{Ceperley_RMP}. The method involves
mapping of the quantum system of particles onto a classical model of
interacting ``ring polymers'', whose elements, ``beads'' or
``time-slices'', are connected by "springs". The method provides a
direct statistical evaluation of quantum canonical averages. In
addition to static properties of the system, dynamical properties
can be also extracted from PIMC simulations.\cite{Ceperley_RMP}

The object of this study is the phonon spectrum, which can be
extracted from the dynamic structure factor, $S({ \bf q},\omega)$.
The definition of $S({ \bf q},\omega)$ in terms of density fluctuations is
\begin{equation}
\label{struct_fun}
S({ \bf q}, \omega) =  \frac{1}{2 \pi n} \int_{-\infty}^{+\infty} dt e^{i\omega t}<\rho_{ { \bf q}}(t) \rho_{-{ \bf q}}(0) >,
\end{equation}
where $\hbar { \bf q}$ and $\hbar \omega$ are the momentum and
energy (we take $\hbar = 1$), $\rho_{ { \bf q}}$ is the
Fourier transform of the density of the solid, and $n$ is the
number density. $S({ \bf q},\omega)$ is usually expressed in terms
of phonons, by writing $S({ \bf q},\omega)$ as a sum of terms
involving the excitation of a single phonon, $S_1({ \bf
q},\omega)$, a pair of phonons, $S_2({ \bf q},\omega)$ and higher
order terms which also include interference between
different terms.~\cite{Glyde,Horner} In most of our simulations we
calculated the $S_1({ \bf q},\omega)$ term. Some calculations of
$S({ \bf q},\omega)$ were also performed, and will be discussed
below. Additional details of the method used to calculate both  $S_1({ \bf q},\omega)$
and $S({ \bf q},\omega)$ and extract phonon spectra
are described in our paper.~\cite{Sorkin}

In the simulations we used samples containing between 128 and 432
atoms. Each atom was represented by a ``ring'' polymer with 64 time slices.
$S({\bf q }, \omega)$ was calculated for
values of $q$ between 0.14 and 1 in relative lattice units
(r.l.u.= 2$\pi/a$, where $a$ is the lattice parameter). The number
density was set to $\rho = 0.02801~(1/\AA^3)$
and the temperature to T=1.6~K. This particular temperature was chosen to compare
the phonon spectrum of $^3$He with those of
$^4$He calculated at the same temperature~\cite{Sorkin}.
A perfect bcc lattice was prepared for the initial
configuration. The effects of Fermi statistics are not taken into account
in our simulations, which is a reasonable approximation for the solid phase
at T=1.6~K.

Statistical errors were estimated by running the PIMC simulations at least
10 times, with different initial conditions in each case.
After each run, $S({\bf q }, \omega)$ ($S_1({\bf q }, \omega)$)
was extracted using the MaxEnt method~\cite{MaxEnt}.
The phonon energy for a given ${\bf q}$ was then calculated by
averaging the positions of the peak of $S({\bf q }, \omega)$ over
the set of the simulation runs. The error bars shown
in the figures below represent the standard deviation.
We collected at least $~$10000 data points in each simulation run
and re-blocked the data~\cite{Sorkin} in blocks of 100-200 points~\cite{Sorkin}.
Each simulation run took about two weeks of 12 Pentium III
PCs running in parallel.

\begin{figure}[ht]
\includegraphics[scale=0.8]{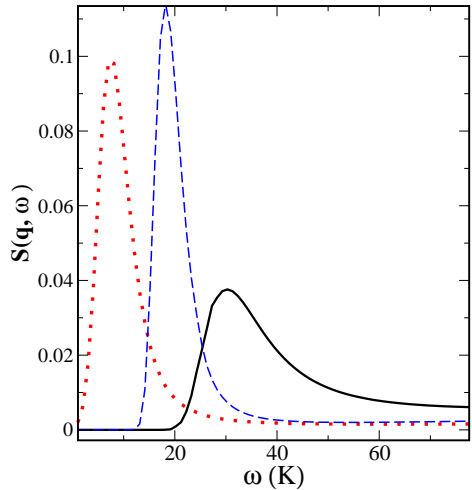}
\caption { \label{S1}  Longitudinal component (solid line)
and transverse components (T$_1$ - dotted line, T$_2$ - dashed line)
of the dynamic structure factor, $S_1({\bf q }, \omega)$,  for q = 0.4 r.l.u. along the [011] direction.
}
\end{figure}

 A typical example of the calculated dynamic
structure factor $S_1({\bf q }, \omega)$  along
the [011] direction is shown in Fig.\ref{S1}. The figure shows both
the longitudinal and transverse phonons. As seen in  Fig.\ref{S1}
the two transverse phonons have narrow and almost symmetric line shapes,
while the longitudinal phonon is quite broad and asymmetric.
This asymmetry exhibits itself in the form of a relatively steep rise
on the low-frequency side and a more gradual decrease on the high-frequency
side. Similar form of the phonon spectral function was obtained by
Koehler and Werthamer~\cite{Koehler}.

\section{Results}

The calculated dispersion relations of longitudinal and transverse phonons,
along the main crystal directions ([001],
[111] and [011]), are shown in Figs.~\ref{l100} - \ref{t111}.
(The numerical values are given in the Appendix).

\begin{figure} [ht]
\includegraphics[scale=0.8]{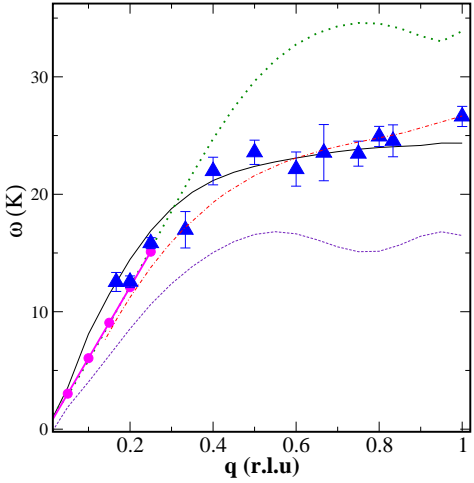}
\caption { \label{l100} Calculated dispersion relation of the
L[001] phonon branch  (triangles) using $S_1({\bf q },\omega)$.
The error bars represent statistical uncertainty.
For reference,  phonon frequencies  calculated by Glyde and Khanna~\cite{Glyde_ph} (solid lines),
Koehler and Werthamer~\cite{Koehler}  (dotted lines) and Horner\cite{Horner_ph}~(dashed line)
are shown. The phonon spectrum at small q (circles) is
estimated using elastic constants measured by Greywall~\cite{Greywall}.
The scaled single-phonon branch of $^4$He (dot-dashed line) is taken from~\cite{Sorkin}. }
\end{figure}
\begin{figure} [ht]
\includegraphics[scale=0.8]{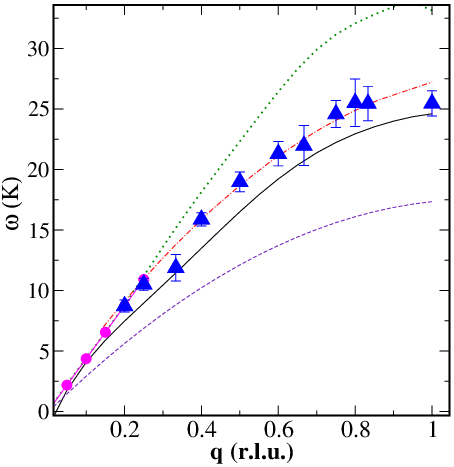}
\caption { \label{t001}Calculated dispersion relation of the
T[001] phonon branch  (triangles) using $S_1({\bf q },\omega)$.
The error bars represent statistical uncertainty.
For reference,  phonon frequencies  calculated by Glyde and Khanna~\cite{Glyde_ph} (solid lines),
Koehler and Werthamer~\cite{Koehler}  (dotted lines) and Horner\cite{Horner_ph}~(dashed line)
are shown. The phonon spectrum at small q (circles) is
estimated using elastic constants measured by Greywall~\cite{Greywall}.
The scaled single-phonon branch of $^4$He (dot-dashed line) is taken from~\cite{Sorkin}.
 }
\end{figure}
\begin{figure} [ht]
\includegraphics[scale=0.8]{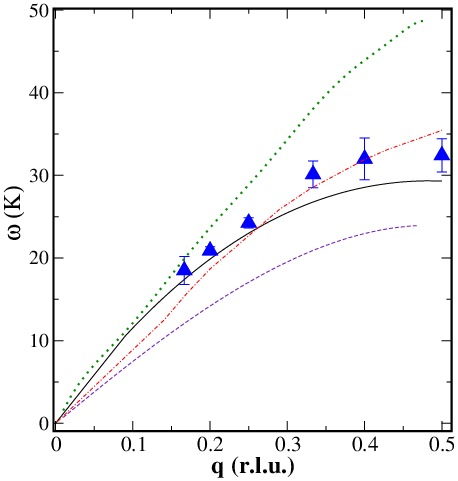}
\caption { \label{l011}Calculated dispersion relation of the
L[011] phonon branch  (triangles) using $S_1({\bf q },\omega)$.
The error bars represent statistical uncertainty.
For reference,  phonon frequencies  calculated by Glyde and Khanna~\cite{Glyde_ph} (solid lines),
Koehler and Werthamer~\cite{Koehler}  (dotted lines) and Horner\cite{Horner_ph}~(dashed line)
are shown.
The scaled single-phonon branch of $^4$He (dot-dashed line) is taken from~\cite{Sorkin}.
}
\end{figure}
\begin{figure} [ht]
\includegraphics[scale=0.8]{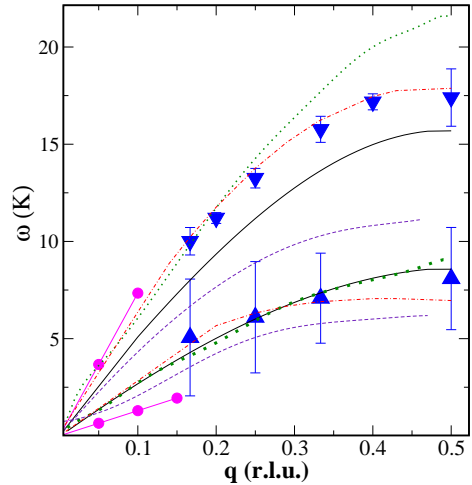}
\caption { \label{t011} Calculated  dispersion relations of
transverse phonon branches along [011] using $S_1({\bf q
},\omega)$. Calculated values are shown for the $T_1$ branch
(triangles down) and $T_2$  branch (triangles up).
The error bars represent statistical uncertainty. For reference,  phonon frequencies
 calculated by Glyde and Khanna~\cite{Glyde_ph} (solid lines),
Koehler and Werthamer~\cite{Koehler}(dotted lines)
and Horner\cite{Horner_ph}~(dashed line)
are shown. The phonon spectra at small q (circles) are
estimated using elastic constants measured by Greywall~\cite{Greywall}.
The scaled single-phonon branches of $^4$He (dot-dashed lines) is taken from~\cite{Sorkin}.
}
\end{figure}
\begin{figure} [ht]
\includegraphics[scale=0.8]{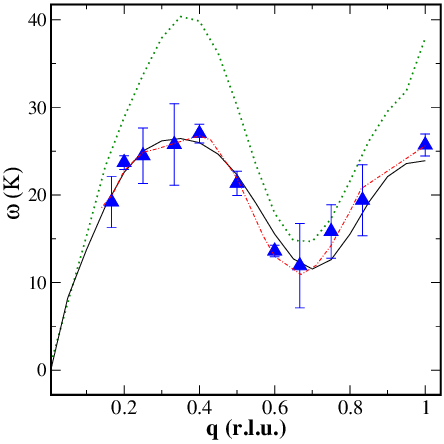}
\caption { \label{l111} Calculated dispersion relation of the
L[111] phonon branch (triangles)
using $S_1({\bf q },\omega)$. The error bars represent statistical uncertainty.
The error bars represent statistical uncertainty. For reference,  phonon frequencies
calculated by Glyde and Khanna~\cite{Glyde_ph} (solid lines) along with frequencies
calculated by Koehler and Werthamer~\cite{Koehler}(dotted lines) are shown.
The scaled single-phonon branch of $^4$He (dot-dashed line) is taken from~\cite{Sorkin}.
}
\end{figure}
\begin{figure} [ht]
\includegraphics[scale=0.8]{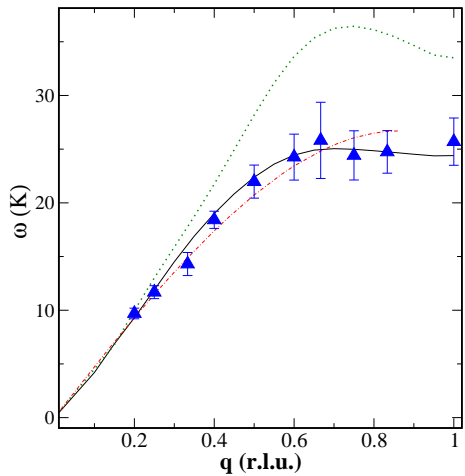}
\caption { \label{t111}  Calculated dispersion relation of the
T[111] phonon branch (triangles) using $S_1({\bf q },\omega)$.
The error bars represent statistical uncertainty.
The error bars represent statistical uncertainty. For reference,  phonon frequencies
calculated by Glyde and Khanna~\cite{Glyde_ph} (solid lines) along with frequencies
calculated by Koehler and Werthamer~\cite{Koehler}(dotted lines) are shown.
The scaled single-phonon branch of $^4$He (dot-dashed line) is taken from~\cite{Sorkin}.
}
\end{figure}

For comparison, we plot the phonon branches calculated at T=0 K
by Glyde and Khanna~\cite{Glyde_ph}, by Koehler and Werthamer~\cite{Koehler}
(at molar volume V=21.5 cm$^3$) and Horner~\cite{Horner_ph}
(at molar volume V=24 cm$^3$). Koehler and Werthamer
suggested that due to the asymmetry of the dynamic structure factor,
the phonon frequency  can be calculated either
at the position of the maximum of $S({\bf q }, \omega)$ or
as the mean of the two half-maxima~\cite{Koehler}. Since we calculated the phonon
spectra using the position of the maximum of $S({\bf q }, \omega)$
only these phonon frequencies were taken from Koehler and Werthamer~\cite{Koehler}.

In addition, we evaluated phonon frequencies in the long wavelength limit (q $\rightarrow$ 0)
using the elastic moduli of bcc $^3$He measured by Greywall\cite{Greywall}
at molar volume V = 21.6 cm$^3$. We also made a fit to  $^4$He phonon branches
calculated by the PIMC method~\cite{Sorkin}, scaling  them by square root of m($^4$He)/m($^3$He),
 the mass ratio of the isotopes, and plot in Figs.~\ref{l100} - \ref{t111}.

Surprisingly, we found that the results of ours simulations
show good agreement with the theoretical calculation of Glyde and Khanne~\cite{Glyde_ph}
(except the phonon branch  $T_1$ [011] in Fig.~\ref{t011}).
In addition, as  can be seen from  Figs.~\ref{l100} - \ref{t111}
the calculated phonon branches of $^3$He
coincide very closely with the scaled branches of $^4$He
(being practically on top of each other for longitudinal phonons).
We conclude that, according to our simulations, the phonon
frequencies of these two isotopes scale as an inverse ratio
of the square root of their atomic masses.

\section{Conclusions}
In conclusion, we calculated the dynamic structure factor for solid $^3$He in the
bcc phase of using PIMC simulations and the MaxEnt method. PIMC
was used to calculate the intermediate scattering function in the
imaginary time from which the dynamic structure factor was
inferred with the MaxEnt method. We extracted both the longitudinal and
transverse phonon branches from the one-phonon dynamic structure
factor. Good agreement between the calculated phonon spectra
and the theoretical prediction of Glyde and Khanne~\cite{Glyde_ph}
has been obtained. The results of our simulations show that the phonon
frequencies of these two isotopes scale as an inverse ratio
of the square root of their atomic masses.

\begin{acknowledgements}
We wish to thank D. Ceperley for many helpful discussions and for
providing us with his  UPI9CD PIMC code. We are grateful to  N. Gov, O.
Pelleg and S. Meni for discussions. This study
was supported in part by the Israel Science Foundation and by the
Technion VPR fund for promotion of research.
\end{acknowledgements}


\section{Appendix: Phonon energies}

The energies of phonons calculated from  $S_1({\bf q},\omega)$ in the bcc phase of $^4$He (molar volume 21 cm$^3$)
and $^3$He (molar volume 21.5 cm$^3$) at T = 1.6 K are listed in Tabs.~\ref{table1}-~\ref{table4}. The phonon energy, $\omega$,
is in units of Kelvin, and the reciprocal lattice vector, $q$, is in relative lattice units
(r.l.u.= 2$\pi/a$, where $a=4.1486\AA$ is the lattice parameter).

\begin{table}[ht]
\caption{\label{table1} Calculated phonon energies, $\omega$(K), of the L[001], T[001], L[111] and T[111] phonon branches 
of bcc $^4$He obtained by using $S_1({\bf q},\omega)$. }

\begin{tabular}{ccccc}
\hline\noalign{\smallskip}
 q (r.l.u) &  L[001] & T[001] &  L[111] & T[111] \\ 
\hline\noalign{\smallskip}
 0.14     &   7.1  $\pm$  1.9     &  6.1 $\pm$   1.0      & 16.2 $\pm$   1.5      &   6.1  $\pm$  1.9    \\ 
 0.20     &   9.1  $\pm$  2.5     &  8.1 $\pm$   1.5      & 20.2 $\pm$   2.8      &   8.1  $\pm$  0.0    \\ 
 0.25     &  12.1  $\pm$  1.5     &  9.1 $\pm$   1.8      & 21.2 $\pm$   2.2      &  10.1  $\pm$  1.8    \\ 
 0.29     &  13.1  $\pm$  1.6     & 10.1 $\pm$   1.5      & 22.2 $\pm$   4.0      &  10.1  $\pm$  1.5    \\ 
 0.33     &  15.5  $\pm$  1.4     & 12.1 $\pm$   1.0      & 22.2 $\pm$   1.2      &  13.1  $\pm$  1.8    \\  
 0.40     &  16.2  $\pm$  1.6     & 13.1 $\pm$   1.1      & 23.2 $\pm$   2.7      &  15.2  $\pm$  2.1    \\ 
 0.43     &  18.2  $\pm$  1.5     & 15.2 $\pm$   1.8      & 23.2 $\pm$   3.4      &  16.2  $\pm$  3.0    \\ 
 0.50     &  19.2  $\pm$  1.1     & 16.2 $\pm$   1.2      & 18.2 $\pm$   1.9      &  18.2  $\pm$  2.2    \\ 
 0.57     &  18.2  $\pm$  1.4     & 18.2 $\pm$   1.5      & 14.2 $\pm$   4.1      &  20.2  $\pm$  2.2    \\ 
 0.60     &  20.2  $\pm$  1.7     & 19.2 $\pm$   0.9      & 11.1 $\pm$   2.0      &  20.2  $\pm$  2.8    \\ 
 0.67     &  21.2  $\pm$  1.7     & 19.2 $\pm$   1.8      &  9.1 $\pm$   2.0      &  22.2  $\pm$  3.4    \\ 
 0.71     &  21.2  $\pm$  1.9     & 20.2 $\pm$   1.9      & 10.1 $\pm$   1.7      &  22.2  $\pm$  4.2    \\ 
 0.75     &  21.2  $\pm$  2.2     & 20.2 $\pm$   1.9      & 13.1 $\pm$   1.7      &  21.2  $\pm$  3.6    \\ 
 0.80     &  21.2  $\pm$  2.2     & 21.2 $\pm$   2.4      & 15.2 $\pm$   1.7      &  22.2  $\pm$  2.3    \\ 
 0.83     &  22.4  $\pm$  1.7     & 22.5 $\pm$   2.3      & 18.2 $\pm$   1.6      &  24.2  $\pm$  4.1    \\ 
 0.86     &  21.2  $\pm$  2.6     & 22.2 $\pm$   2.5      & 22.2 $\pm$   1.8      &  23.2  $\pm$  3.1    \\
\hline\noalign{\smallskip}
\end{tabular}
\end{table}

\begin{table}
\caption{\label{tab:table1} Calculated phonon energies $\omega$(K) of the L[011], T$_2$[011] and  T$_1$[011] phonon branches 
of bcc $^4$He obtained by using $S_1({\bf q},\omega)$. }
\begin{tabular}{ccccc}
\hline\noalign{\smallskip}
q (r.l.u) &  L[011]  &  T$_1$[011]  &  T$_2$[011] \\ 
\hline\noalign{\smallskip}
 0.20 &  16.2  $\pm$  2.0 &  5.1 $\pm$   1.2  & 10.1 $\pm$   1.3  \\ 
 0.25 &  33.3  $\pm$  2.3 &  5.1  $\pm$   1.7  & 12.1 $\pm$   1.3  \\ 
 0.29 &  21.2  $\pm$  2.7 &  6.0 $\pm$   1.7  & 12.1 $\pm$   1.0  \\
 0.33 &  26.3  $\pm$  1.5 &  6.05 $\pm$   1.1  & 14.2 $\pm$   2.0  \\ 
 0.40 &  29.3  $\pm$  1.9 &  6.11 $\pm$   1.0  & 15.2 $\pm$   1.5  \\
 0.43 &  26.3  $\pm$  2.8 &  6.1 $\pm$   1.6  & 16.2 $\pm$   2.5  \\ 
 0.50 &  31.3  $\pm$  2.5 &  6.12 $\pm$   2.4  & 15.2 $\pm$   1.8 \\
\hline\noalign{\smallskip}
\end{tabular}
\end{table}

\begin{table}
\caption{\label{tab:table1} Calculated phonon energies $\omega$(K) of the L[001], T[001], L[111] and T[111] phonon branches of bcc $^3$He
obtained by  using $S_1({\bf q},\omega)$. }
\begin{tabular}{cccccc}
\hline\noalign{\smallskip}
 q (r.l.u) &  L[001] & T[001] &  L[111] & T[111] \\  
\hline\noalign{\smallskip}
 0.20     &  12.6  $\pm$  0.5     &  8.7 $\pm$   0.5      & 19.2 $\pm$   2.9      &  9.7  $\pm$  0.4    \\   
 0.25     &  15.8  $\pm$  0.5     & 10.5 $\pm$   0.5      & 24.5 $\pm$   3.1      &  11.7  $\pm$  0.6    \\   
 0.33     &  17.0  $\pm$  1.6     & 11.9 $\pm$   1.1      & 23.9 $\pm$   3.9      &  14.3  $\pm$  1.7    \\   
 0.40     &  22.0  $\pm$  1.2     & 15.9 $\pm$   0.5      & 25.8 $\pm$   4.7      &  18.4  $\pm$  0.8    \\   
 0.50     &  23.6  $\pm$  0.6     & 18.9 $\pm$   0.9      & 21.3 $\pm$   1.1      &  21.9  $\pm$  1.5    \\   
 0.60     &  22.1  $\pm$  1.5     & 21.3 $\pm$   1.0      & 13.5 $\pm$   1.5      &  24.7  $\pm$  2.1    \\   
 0.67     &  23.5  $\pm$  2.4     & 22.0 $\pm$   1.7      & 11.9 $\pm$   0.7      &  25.8  $\pm$  3.5                     \\   
 0.75     &  23.5  $\pm$  1.6     & 24.5 $\pm$   1.1      & 15.8 $\pm$   3.1      &  24.3  $\pm$  2.1    \\   
 0.80     &  24.9  $\pm$  0.9     & 25.5 $\pm$   2.0      & 15.0 $\pm$   3.6      &  24.7  $\pm$  1.9    \\   
 0.83     &  24.6  $\pm$  1.4     & 25.2 $\pm$   1.2      & 19.4 $\pm$   4.1      &  25.2  $\pm$  2.1    \\   
 1.00     &  26.6  $\pm$  0.8     & 25.4 $\pm$   1.3      & 25.7 $\pm$   0.7      &  25.7  $\pm$  2.7    \\
\hline\noalign{\smallskip}
\end{tabular}
\end{table}

\begin{table}
\caption{\label{table4} Calculated phonon energies $\omega$(K) of the L[011], T$_2$[011] and  T$_1$[011] phonon branches  of $^3$He
obtained by using $S_1({\bf q},\omega)$. }
\begin{tabular}{cccc}
\hline\noalign{\smallskip}
q (r.l.u) &  L[011]  &  T$_1$[011]  &  T$_2$[011] \\ 
\hline\noalign{\smallskip}
 0.17 &  18.5  $\pm$  1.7 &  1.8 $\pm$   1.5  & 10.0 $\pm$   0.7  \\  
 0.20 &  20.9  $\pm$  0.5 &  2.8 $\pm$   3.8  & 11.2 $\pm$   0.3  \\  
 0.25 &  24.2  $\pm$  1.0 &  6.1 $\pm$   2.8  & 13.1 $\pm$   0.6  \\  
 0.33 &  30.1  $\pm$  1.6 &  7.1 $\pm$   0.5  & 15.8 $\pm$   0.7  \\  
 0.40 &  32.0  $\pm$  2.5 &  7.1 $\pm$   0.0  & 17.2 $\pm$   0.4  \\  
 0.50 &  32.4  $\pm$  2.0 &  8.1 $\pm$   2.5  & 17.6 $\pm$   1.1  \\
\hline\noalign{\smallskip}
\end{tabular}
\end{table}

\end{document}